\documentclass[12pt,letterpaper]{article}
\usepackage{natbib,epsfig,graphicx,setspace}
\usepackage{amsmath,amsthm,amssymb}
\usepackage{amsmath,enumerate,mathrsfs}
\bibpunct{(}{)}{;}{a}{,}{,}

\def\bg{\begin{figure}[tpbh]\begin{center}}
\def\eg{\end{center}\end{figure}}

\numberwithin{equation}{section} 

\def\bsigma{{\mbox{\boldmath $\sigma$}}}
\def\btheta{{\mbox{\boldmath $\theta$}}}
\def\bbeta{{\mbox{\boldmath $\beta$}}}
\def\balpha{{\mbox{\boldmath $\alpha$}}}
\def\bpi{{\mbox{\boldmath $\pi$}}}

\def\bS{{\mbox{\boldmath $S$}}}
\def\bm{{\mbox{\boldmath $m$}}}
\def\bx{{\mbox{\boldmath $x$}}}

\def\bQ{{\mbox{\boldmath $Q$}}}
\def\bR{{\mbox{\boldmath $R$}}}
\def\b0{{\mbox{\boldmath $0$}}}
\def\by{{\mbox{\boldmath $y$}}}
\def\bta{{\mbox{\boldmath $\eta$}}}
\def\blambda{{\mbox{\boldmath $\lambda$}}}

\newcommand{\ann}{\emph{Annals of Statistics}}

\newcommand{\jasa}{\emph{Journal of the American Statistical Association}}

\newtheorem{thm}{Theorem}[section]

\newtheorem{alg}{Algorithm}[section]

\begin{document}

\title{Semiparametric Mixtures of Regressions with Single-index for Model Based Clustering}
\author{ Sijia Xiang\thanks{School of Data Sciences, Zhejiang University of Finance and Economics. E-mail:
sjxiang@zufe.edu.cn. Xiang's research is supported by Zhejiang Provincial NSF of China grant LQ16A010002 and NSF of China grant 11601477.} and Weixin Yao\thanks{ Department of Statistics,  University of California, Riverside, California,
U.S.A. E-mail: weixin.yao@ucr.edu. Yao's research is supported by NSF grant DMS-1461677 and Department of Energy Award No: 10006272.}}

\date{}
\maketitle{}

\begin{abstract}
In this article, we propose two classes of semiparametric mixture regression models with single-index for model based clustering. Unlike many semiparametric/nonparametric mixture regression models that can only be applied to low dimensional predictors, the new semiparametric models can easily incorporate high dimensional predictors into the nonparametric components. The proposed models are very general, and many of the recently proposed semiparametric/nonparametric mixture regression models are indeed special cases of the new models. Backfitting estimates and the corresponding modified EM algorithms are proposed to achieve optimal convergence rates for both parametric and nonparametric parts. We establish the identifiability results of the proposed two models and investigate the asymptotic properties of the proposed estimation procedures. Simulation studies are conducted to demonstrate the finite sample performance of the proposed models. An application of NBA data by new models reveals some new findings.
\end{abstract}


\section{Introduction}

Mixtures of regression models are commonly used as model based clustering methods to reveal the  relationship among interested variables if the whole population is inhomogeneous and consists of several homogeneous subgroups. They have been widely used in many areas such as econometrics, biology, and epidemiology. For a general account of traditional parametric mixture models, please see, for example, \citet{lindsay95}, \citet{bohning99}, \citet{mcLachlan:et00}, and \citet{fruhwirth06}. However, the traditional mixture of regression models requires strong parametric assumption: liner component regression functions, constant component variance, and constant component proportions. The fully parametric hierarchical mixtures of experts model \citep{jordan94} has been proposed to allow the component proportions to depend on the covariates in machine learning. Recently, many semiparametric and nonparametric mixture regression models have been proposed to relax the parametric assumption of mixture regression models. See, for example, \citet{young2010, huang2012, cao2012, huang2013, huang2014}, among others. However, most of those existing semparametirc or nonparametric mixture regressions can only be applied for low dimensional predictors due to ``curse of dimensionality". It will be desirable to be able to relax parametric assumptions of traditional mixtures of regression models when the dimension of predictors is high.

In this article, we propose a mixture of single-index models (MSIM) and a mixture of regression models with varying single-index proportions (MRSIP) to reduce the dimension of high dimensional predictors before modeling them nonparametrically. Many existing popular models can be considered as special cases of the proposed two models. \citet{huang2013} proposed the nonparametric mixture of regression models
\[Y|_{X=x}\sim \sum_{j=1}^k \pi_j(x)\phi(Y_i|m_j(x),\sigma_j^2(x)),\]
 where $\pi_j(x), m_j(x),$ and $\sigma_j^2(x)$ are unknown smoothing functions, and $\phi(y|\mu,\sigma^2)$ is the normal density with mean $\mu$ and variance $\sigma^2$. Their proposed model can drastically reduce the modelling bias when the strong parametric assumption of traditional mixture of linear regression models does not hold. However, the above model is not applicable to high dimensional predictors due to the kernel estimation used for nonparametric parts. To solve the above problem, we propose a \emph{mixture of single-index models}
 \begin{equation}
Y|_{\bx}\sim\sum_{j=1}^k\pi_j(\balpha^T\bx)\phi(Y_i|m_j(\balpha^T\bx),\sigma_j^2(\balpha^T\bx)),
\label{p3-model}
\end{equation}
in which the single index $\balpha^T\bx$ transfers the high dimensional nonparametric problem to a univariate nonparametric problem. When $k=1$, model (\ref{p3-model}) reduces to a single index model \citep{ichiura1993, hardle1993}. If $\bx$ is a scalar, then model (\ref{p3-model}) reduces to the nonparametric mixture of regression model proposed by \citet{huang2013}. \citet{zeng12} also applied the single index idea to the component means and variance and assumed that component proportions do not depend on the predictor $\bx$. However, \citet{zeng12} did not give any theoretical properties of their proposed estimates.  % The new model can naturally incorporate nonparametric effects of the high dimensional predictors and relax the traditional parametric assumption of mixture regression models.

\citet{young2010} and \citet{huang2012} proposed a semiparametric mixture of regression models
\[Y|_{X=x}\sim \sum_{j=1}^k \pi_j(\bx)\phi(Y_i|\bx^T\bbeta_j,\sigma_j^2),\]
where  $\pi_j(\bx)$ is an unknown smoothing function, to combine nice properties of both nonparametric mixture regression models and traditional parametric mixture regression models. Their semiparametric mixture models assume that component proportions depend on covariates nonparametrically to reduce the modelling bias while component regression functions are still assumed to be linear to have better model interpretation. However, their estimation procedures cannot be applied if the dimension of predictors $\bx$ is high due to kernel estimation used for $\pi_j(\bx)$. We propose a \emph{mixture of regression models with varying single-index proportions}
\begin{equation}
Y|_{X=x}\sim \sum_{j=1}^k \pi_j(\balpha^T\bx)\phi(Y_i|\bx^T\bbeta_j,\sigma_j^2),
\label{p41-model}
\end{equation}
which uses the idea of single index to model the nonparametric effect of predictors on component proportions, while allowing easy interpretation of linear component regression functions. When $k=1$, model (\ref{p41-model}) reduces to the traditional linear regression model. If $\bx$ is a scalar, then model (\ref{p41-model}) reduces to the semiparametric mixture models considered by \citet{young2010} and \citet{huang2012}. Modeling component proportions nonparametrically can reduce the modelling bias and better cluster the data when the traditional parametric assumptions of component proportions do not hold \citep{young2010, huang2012}.

We prove the identifiability results of proposed two models under some mild conditions. We propose a modified EM algorithm by combining the ideas of backfitting algorithm, kernel estimation, and local likelihood to estimate global parameters and nonparametric functions. In addition, the asymptotic properties of the proposed estimation procedures are also investigated. Simulation studies are conducted to demonstrate the finite sample performance of the proposed models. An application of NBA data by new models reveals some new interesting findings.

The rest of the paper is organized as follows. In Section \ref{sec:p3-s2}, we introduce the MSIM and study its identifiability result. A one-step and a fully-iterated backfitting estimate are proposed, and their asymptotic properties are also studied. In Section \ref{sec:p3-s3}, we introduce the MRSIP. The identifiability result and asymptotic properties of the proposed estimates are given. In Section \ref{sec:p3-s4} and Section \ref{realdata}, we use Monte Carlo studies and a real data example to demonstrate the finite sample performance of the proposed two models. A discussion section is given in Section \ref{discussion} and we defer the technical conditions and proofs in the supplemental material.

\section{Mixtures of Single-index Models}
\label{sec:p3-s2}
\subsection{Model Definition and Identifiability}
Assume that $\{(\bx_i,Y_i),i=1,...,n\}$ is a random sample from the population $(\bx,Y)$, where $\bx$ is $p$-dimensional and $Y$ is univariate. Let $\mathcal{C}$ be a latent variable, and has a discrete distribution $P(\mathcal{C}=j|\bx)=\pi_j(\balpha^T\bx)$ for $j=1,...,k$. Conditional on $\mathcal{C}=j$ and $\bx$, $Y$ follows a normal distribution with mean $m_j(\balpha^T\bx)$ and variance $\sigma_j^2(\balpha^T\bx)$. Without observing $\mathcal{C}$, the conditional distribution of $Y$ given $\bx$ can be written as:
\begin{equation*}
Y|_{\bx}\sim\sum_{j=1}^k\pi_j(\balpha^T\bx)\phi(Y_i|m_j(\balpha^T\bx),\sigma_j^2(\balpha^T\bx)).
%\label{p3-model}
\end{equation*}
The above model is the proposed mixture of single-index models. Throughout the paper, we assume that $k$ is fixed, and refer to model (\ref{p3-model}) as a finite semiparametric mixture of regression models, since $\pi_j(\cdot)$, $m_j(\cdot)$ and $\sigma_j^2(\cdot)$ are all nonparametric. In the model (\ref{p3-model}), we use the same index $\balpha$ for all components. But our proposed estimation procedure and asymptotic results can be easily extended to the cases where components have different index $\balpha$.
%\end{document}
%Many existing popular models can be considered special cases of the proposed model (\ref{p3-model}) if the  different. When $k=1$ and $\pi_j(\cdot)$ and $\sigma_j^2(\cdot)$ are constant, model (\ref{p3-model}) reduces to a single index model \citep{ichiura1993, hardle1993}. If $\pi_j(\cdot)$ and $\sigma_j^2(\cdot)$ are constant, and $m_j(\cdot)$ are identity functions, then model (\ref{p3-model}) reduces to a finite mixture of linear regression models \citep{goldfeld1973}. If $\bx$ is a scalar, then model (\ref{p3-model}) reduces to the nonparametric mixture of regression model proposed by \citet{huang2013}. %Therefore,  many existing popular models can be considered as special cases of the proposed model (\ref{p3-model}).

Compared to \citet{huang2013}, the appeal of the proposed MSIM is that by using an index $\balpha^T\bx$, the so-called ``curse of dimensionality'' in fitting multivariate nonparametric regression functions is avoided. It is of dimension-reduction structure in the sense that, given the estimate of $\balpha$, denoted by $\hat{\balpha}$, we can use the univariate $\hat{\balpha}^T\bx$ as the covariate and simplify the model (\ref{p3-model}) to the nonparametric mixture regression model proposed by \citet{huang2013}. Therefore, model (\ref{p3-model}) is a reasonable compromise between fully parametric and fully nonparametric modeling.

%Xiang and Yao (2013) studied a semiparametric mixture of regression models with component mean functions being smooth functions of a covariate. They assume that given $X=x$, the distribution of $Y$ can be written as $Y|_{X=x}\sim \sum_{j=1}^k \pi_j\phi(Y_i|m_j(x),\sigma_j^2)$, where $X$ is assumed to be univariate, due to the ``curse of dimensionality". Compared to it, model (\ref{p3-model}) has remarkably wider application, since it allows for multivariate predictors.

Identifiability is a major concern for most mixture models. Some well known identifiability results of finite mixture models include: mixture of univariate normals is identifiable up to relabeling \citep{titterington1985} and finite mixture of regression models is identifiable up to relabeling provided that covariates have a certain level of variability \citep{henning2000}. The following theorem establishes the identifiability result of the model (\ref{p3-model}) and its proof is given in the supplemental material. %Section \ref{p3-sec:proofs}.

%\newpage

\begin{thm}
\label{p3-thm:iden}
Assume that
\begin{enumerate}%[(i)]
\item $\pi_j(z)$, $m_j(z)$, and $\sigma_j^2(z)$ are differentiable and not constant on the support of $\balpha^T\bx$, $j=1,...,k$;
 \item The $\bx$ are continuously distributed random variables that have a joint probability density function;
  \item The support of $\bx$ is not contained in any proper linear subspace of $\mathbb{R}^p$;
 \item  $\|\balpha\|=1$ and the first nonzero element of $\balpha$ is positive;
 \item For any $1\le i\ne j\le k$, \[\sum_{l=0}^{1}
\|m^{(l)}_i(z)-m^{(l)}_j(z)\|^2+
\sum_{l=0}^{1}\|\sigma^{(l)}_i(z)-\sigma^{(l)}_j(z)\|^2\ne 0,\] for any $z$ where
$g^{(l)}$ is the $l$th derivative of $g$ and equal to $g$ if $l=0$.
\end{enumerate}
Then, model (\ref{p3-model}) is identifiable.
\end{thm}
%The above condition transversality of two smooth curves (Huang et al., 2013) implies that the mean and variance functions of any two components cannot be tangent to each other.

\subsection{Estimation Procedure}
\label{sec:p3-s3}
In this subsection, we propose a one-step estimation procedure and a backfitting algorithm to estimate the nonparametric functions and the single index of the model (\ref{p3-model}).

Let $\ell^{*(1)}(\bpi,\bm,\bsigma^2,\balpha)$ be the log-likelihood of the collected data $\{(\bx_i,Y_i),i=1,...,n\}$ from the model (\ref{p3-model}). That is:
\begin{equation}
\ell^{*(1)}(\bpi,\bm,\bsigma^2,\balpha)=\sum_{i=1}^n\log\{\sum_{j=1}^k\pi_j(\balpha^T\bx_i)\phi(Y_i|m_j(\balpha^T\bx_i),\sigma_j^2(\balpha^T\bx_i))\},
\label{p3-lstar}
\end{equation}
where $\bpi(\cdot)=\{\pi_1(\cdot),...,\pi_{k-1}(\cdot)\}^T$, $\bm(\cdot)=\{m_1(\cdot),...,m_k(\cdot)\}^T$, and $\bsigma^2(\cdot)=\{\sigma_1^2(\cdot),...,\sigma_k^2(\cdot)\}^T$. Since $\bpi(\cdot)$, $\bm(\cdot)$ and $\bsigma^2(\cdot)$ consist of nonparametric functions, (\ref{p3-lstar}) is not ready for maximization. %We propose a one-step backfitting procedure to estimate the nonparametric functions and the index parameters.

Note that for the model (\ref{p3-model}), the space spanned by the single index $\balpha$ is in fact the central mean subspace of $Y|\bx$ \citep{cook02} in the literature of sufficient dimension reduction. Therefore, we can employ existing sufficient dimension reduction methods to find an initial estimate of $\balpha$. Please see, for example, \citet{li1991,lzc2005,wang08,luo09, ma12a,ma12b}. In this article, we will simply employ sliced inverse regression \citep{li1991} to obtain an initial estimate of $\balpha$, denoted by $\tilde{\balpha}$.

Given the estimated single index $\tilde{\balpha}$, the nonparametric functions $\bpi(z)$, $\bm(z)$ and $\bsigma^2(z)$ can then be estimated by maximizing the following local log-likelihood function:
\begin{equation}
\ell^{(1)}_1(\bpi,\bm,\bsigma^2)=\sum_{i=1}^n\log\{\sum_{j=1}^k\pi_j(\tilde{\balpha}^T\bx_i)\phi(Y_i|m_j(\tilde{\balpha}^T\bx_i),\sigma_j^2(\tilde{\balpha}^T\bx_i))\}K_h(\tilde{\balpha}^T\bx_i-z),
\label{p3-l1}
\end{equation}
where $K_h(z)=\frac{1}{h}K(\frac{z}{h})$, $K(\cdot)$ is a kernel density function, and $h$ is a tuning parameter. Let $\hat{\bpi}(\cdot)$, $\hat{\bm}(\cdot)$ and $\hat{\bsigma}^2(\cdot)$ be the estimates that maximize (\ref{p3-l1}). The above estimates are the proposed \emph{one-step estimate}.

We propose a modified EM-type algorithm to maximize $\ell_1^{(1)}$. In practice, we usually want to evaluate unknown functions at a set of grid points, which in this case, requires us to maximize local log-likelihood functions at a set of grid points. If we simply employ the EM algorithm separately for each grid point, the labels in the EM algorithm may change at different grid points, and we may not be able to get smoothed estimated curves \citep{huang2012}. Therefore, we propose the following modified EM-type algorithm, which estimates the nonparametric functions simultaneously at a set of grid points, say $\{u_t,t=1,...,N\}$, and provides a unified label of each observation across all grid points. %Let  be a set of grid points where the unknown functions are evaluated, and $N$ be the number of grid points.\\

\begin{alg}
Modified EM-type algorithm to maximize (\ref{p3-l1}) given the single index estimate $\tilde{\balpha}$.
\begin{description}
\item[E-step:] Calculate the expectations of component labels based on estimates from $l^{th}$ iteration:
\begin{equation}
p_{ij}^{(l+1)}=\frac{\pi_j^{(l)}(\tilde{\balpha}^T\bx_i)\phi(Y_i|m_j^{(l)}(\tilde{\balpha}^T\bx_i),\sigma_j^{2(l)}(\tilde{\balpha}^T\bx_i))}{\sum_{j=1}^k
\pi_j^{(l)}(\tilde{\balpha}^T\bx_i)\phi(Y_i|m_j^{(l)}(\tilde{\balpha}^T\bx_i),\sigma_j^{2(l)}(\tilde{\balpha}^T\bx_i))},
\label{p3-step1e}
\end{equation}
where $i=1,\ldots,n,j=1,\ldots,k.$
\item[M-step:] Update the estimates
\begin{align}
&\pi_j^{(l+1)}(z)=\frac{\sum_{i=1}^np_{ij}^{(l+1)}K_h(\tilde{\balpha}^T\bx_i-z)}{\sum_{i=1}^nK_h(\tilde{\balpha}^T\bx_i-z)}\label{p3-step1m1},\\
&m_j^{(l+1)}(z)=\frac{\sum_{i=1}^np_{ij}^{(l+1)}Y_iK_h(\tilde{\balpha}^T\bx_i-z)}{\sum_{i=1}^np_{ij}^{(l+1)}K_h(\tilde{\balpha}^T\bx_i-z)}\label{p3-step1m2},\\
&\sigma_j^{2(l+1)}(z)=\frac{\sum_{i=1}^np_{ij}^{(l+1)}(Y_i-m_j^{(l+1)}(z))^2K_h(\tilde{\balpha}^T\bx_i-z)}{\sum_{i=1}^np_{ij}^{(l+1)}K_h(\tilde{\balpha}^T\bx_i-z)},\label{p3-step1m3}
\end{align}
for $z\in\{u_t,t=1,...,N\}$ and $j=1,\ldots,k$. We then update $\pi_j^{(l+1)}(\tilde{\balpha}^T\bx_i)$, $m_j^{(l+1)}(\tilde{\balpha}^T\bx_i)$ and $\sigma_j^{2(l+1)}(\tilde{\balpha}^T\bx_i)$, $i=1,...,n$, by linear interpolating $\pi_j^{(l+1)}(u_t)$, $m_j^{(l+1)}(u_t)$ and $\sigma_j^{2(l+1)}(u_t)$, $t=1,...,N$, respectively.
\end{description}
\end{alg}
Note that in the M-step, the nonparametric functions are estimated simultaneously at a set of grid points, and therefore, the classification probabilities in the the E-step can be estimated globally to avoid the label switching problem \citep{yao2009}. If the sample size $n$ is not too large, one can also take all $\{\tilde{\balpha}^T\bx_i, i=1,\ldots,n\}$ as grid points for $z$ in the M-step.

The initial estimate $\tilde{\balpha}$ by SIR does not make use of the mixture information and thus is not efficient. Given one step estimate $\hat{\bpi}(\cdot)$, $\hat{\bm}(\cdot)$ and $\hat{\bsigma}^2(\cdot)$, we can further improve the estimate of $\balpha$ by maximizing
\begin{equation}
\ell^{(1)}_2(\balpha)=\sum_{i=1}^n\log\{\sum_{j=1}^k\hat{\pi}_j(\balpha^T\bx_i)\phi(Y_i|\hat{m}_j(\balpha^T\bx_i),\hat{\sigma}_j^2(\balpha^T\bx_i))\},
\label{p3-l2}
\end{equation}
with respect to $\balpha$. The proposed \emph{fully iterative backfitting estimator} of $\balpha$, denoted by $\hat{\balpha}$, iterates the above two steps until convergence.
%\subsubsection{Computing Algorithm}
%\label{sec:comp}
%We now describe two proposed algorithms to estimate the model (\ref{p3-model}) .
%\noindent\textbf{One-step Estimator (OS)}\\
%\begin{description}
%\item[Step 1:] Obtain an estimate of the index parameter $\balpha$.\\
%Apply sliced inverse regression (Li, 1991) to obtain the estimate of $\balpha$, denoted by $\hat{\balpha}$.
%\item[Step 2:]
\begin{alg}
Fully iterative backfitting estimator (FIB)
%To improve the estimation efficiency, we propose the following \emph{fully iterative backfitting estimator}.\\
\begin{description}
\item[Step 1:] %Obtain an initial estimate of the index parameter $\balpha$.\\
Apply sliced inverse regression (SIR) to obtain an initial estimate of the single index parameter $\balpha$, denoted by $\tilde{\balpha}$.
\item[Step 2:] %Modified EM-type algorithm to maximize $\ell^{(1)}_1$ in (\ref{p3-l1}).\\
Given $\tilde{\balpha}$, apply the modified EM-algorithm (\ref{p3-step1e})|(\ref{p3-step1m3}) to maximize $\ell^{(1)}_1$ in (\ref{p3-l1}) to obtain the estimates $\hat{\bpi}(\cdot)$, $\hat{\bm}(\cdot)$, and $\hat{\bsigma}^2(\cdot)$.
\item[Step 3:] Given $\hat{\bpi}(\cdot)$, $\hat{\bm}(\cdot)$, and $\hat{\bsigma}^2(\cdot)$ from Step 2, update the estimate of $\balpha$ by maximizing $\ell^{(1)}_2$ in (\ref{p3-l2}).
\item[Step 4:] Iterate Steps 2 - 3 until convergence.
%Given $\hat{\bpi}(\cdot)$, $\hat{\bm}(\cdot)$, and $\hat{\bsigma}^2(\cdot)$ from Step 2, one can update the estimate of $\balpha$, denoted by $\hat{\balpha}$, which maximizes $\ell^{(1)}_2$ defined in (\ref{p3-l2}) using some numerical methods, and iterate the procedure to further improve the efficiency. %The fully iterative backfitting estimator is at least as efficient as the one-step estimator, but the one-step estimator achieves the same efficiency in some important applications with added computational convenience.
\end{description}
\end{alg}
%Our simulation studies in Section \ref{sec:p3-s4} demonstrate that FIB provides better estimate of $\balpha$ than SIR.

\subsection{Asymptotic Properties}
\label{sec:asympmodel1}
The asymptotic properties of the proposed estimates are investigated below. Let $\btheta(z)=(\bpi^T(z),\bm^T(z),(\bsigma^2)^T(z))^T$. Define \begin{align*}
\ell(\btheta(z),y)&=\log\sum_{j=1}^k\pi_j(z)\phi\{y|m_j(z),\sigma_j^2(z)\},\\
q_1(z)&=\frac{\partial\ell(\btheta(z),y)}{\partial\theta},\\
q_2(z)&=\frac{\partial^2\ell(\btheta(z),y)}{\partial\theta\partial\theta^T},\\
\mathcal{I}^{(1)}_\theta(z)&=-E[q_2(Z)|Z=z],\\
\Lambda_1(u|z)&=E[q_1(z)|Z=u].
\end{align*}

Under further conditions defined in the supplemental material, the asymptotic properties of the one-step estimates $\hat{\bpi}(\cdot)$, $\hat{\bm}(\cdot)$, and $\hat{\bsigma}^2(\cdot)$ are given in the following theorem.
\begin{thm}
\label{p3-thm:nonp}
Assume that conditions (C1)-(C7) in the supplemental material hold. Then, as $n\rightarrow\infty$, $h\rightarrow 0$ and $nh\rightarrow\infty$, we have
\begin{align}
\sqrt{nh}\{\hat{\btheta}(z)-\btheta(z)-\mathcal{B}_1+o_p(h^2)\}\overset{D}{\rightarrow} N\{0,\nu_0f^{-1}(z)\mathcal{I}^{(1)}_\theta(z)\},
\end{align}
where \[\mathcal{B}_1(z)=\mathcal{I}^{(1)-1}_\theta\left\{\frac{f'(z)\Lambda_1^{'}(z|z)}{f(z)}+\frac{1}{2}\Lambda_1^{''}(z|z)\right\}\kappa_2h^2,\] with $f(\cdot)$ the marginal density function of $\balpha^T\bx$, $\kappa_l=\int t^lK(t)dt$ and $\nu_l=\int t^lK^2(t)dt$.
\end{thm}
%\noindent $\textbf{Remark 1.}$ The fully iterative backfitting estimator is at least as efficient as the one-step estimator, but the one-step estimator achieves the same efficiency in some important applications with added computational convenience.
Note that the asymptotic variance of $\hat{\btheta}(z)$ is the same as those given in \citet{huang2013}. Thus, the nonparametric functions can be estimated with the same accuracy  as it would have if the single index $\balpha^T\bx$ were known. This is expected since the single index $\balpha$ can be estimated at a root $n$ convergence rate which is much faster than $\hat{\btheta}(z)$. In addition, note that the one-step estimates of $\btheta(z)$ have the same asymptotic variance (up to the first order) as the full iterative backfitting algorithm but with much less computations. Our simulation results in Section \ref{sec:p3-s4} further confirm this result.

The next theorem gives the asymptotic results of the $\hat{\balpha}$ given by full iterative backfitting algorithm.
\begin{thm}
\label{p3-thm:index}
Assume that conditions (C1)-(C8) in the supplemental material hold. Then, as $n\rightarrow\infty$, $nh^4\rightarrow 0$, and $nh^2/\log(1/h)\rightarrow\infty$,
\begin{equation}
\sqrt{n}(\hat{\balpha}-\balpha)\overset{D}{\rightarrow} N(0,\bQ_1^{-1}),
\end{equation}
where \[\bQ_1=E\left[\{\bx\btheta'(Z)\}q_2(Z)\{\bx\btheta'(Z)\}^T-\bx\btheta'(Z)q_2(Z)\mathcal{I}^{(1)-1}_\theta(Z)E\{q_2(Z)[\bx\btheta'
(Z)]^T|Z\}\right].\]
\end{thm}

%\newpage

\section{Mixtures of Regression Models with Varying Single-Index Proportions }
\label{sec:p3-s3}
\subsection{Model Definition and Identifiability}
The MRSIP assumes that $P(\mathcal{C}=j|\bx)=\pi_j(\balpha^T\bx)$ for $j=1,...,k$, and conditional on $\mathcal{C}=j$ and $\bx$, $Y$ follows a normal distribution with mean $\bx^T\bbeta_j$ and variance $\sigma^2_j$. That is,
\begin{equation*}
Y|_{\bx}\sim\sum_{j=1}^k\pi_j(\balpha^T\bx)N(\bx^T\bbeta_j,\sigma_j^2).
%\label{p41-model}
\end{equation*}
Since $\pi_j(\cdot)$'s are nonparametric, model (\ref{p41-model}) is also a finite semiparametric mixture of regression models. The linear component regression functions $\bx^T\bbeta_j$ enjoy simple interpretation, while nonparametric functions $\pi_j(\balpha^T\bx)$
can incorporate the effects of predictors on component proportions more flexibly to reduce the modeling bias. See \citet{young2010,huang2013} for more information. We first prove the identifiability result of the model (\ref{p41-model}) in the following theorem and its proof is given in the supplemental material.

\begin{thm}
\label{p4-thm:iden}
Assume that
\begin{enumerate}%[(i)]
 \item $\pi_j(z)>0$ are differentiable and not constant on the support of $\balpha^T\bx$, $j=1,...,k$;
  \item The component of $\bx$ are continuously distributed random variables that have a joint probability density function;
  \item The support of $\bx$ contains an open set in $\mathbb{R}^p$ and is not contained in any proper linear subspace of $\mathbb{R}^p$;
  \item $\|\balpha\|=1$ and the first nonzero element of $\balpha$ is positive;
  \item $(\bbeta_j,\sigma_j^2)$, $j=1,...,k$, are distinct pairs.
\end{enumerate}
Then, model (\ref{p41-model}) is identifiable.
\end{thm}

\subsection{Estimation Procedure}
The log-likelihood of the collected data for the model (\ref{p41-model}) is:
\begin{equation}
\ell^{*(2)}(\bpi,\bsigma^2,\balpha,\bbeta)=\sum_{i=1}^n\log\{\sum_{j=1}^k\pi_j(\balpha^T\bx_i)\phi(Y_i|\bx_i^T\bbeta_j,\sigma_j^2)\},
\label{p41-lstar}
\end{equation}
where $\bpi(\cdot)=\{\pi_1(\cdot),...,\pi_{k-1}(\cdot)\}^T$, $\bsigma^2=\{\sigma_1^2,...,\sigma_k^2\}^T$, and $\bbeta=\{\bbeta_1,...,\bbeta_k\}^T$. Since $\bpi(\cdot)$ consists of nonparametric functions, (\ref{p41-lstar}) is not ready for maximization. We propose a backfitting algorithm to iterate between estimating the parameters $(\balpha,\bbeta,\bsigma^2)$ and the nonparametric functions $\bpi(\cdot)$.

Given the estimates of $(\balpha,\bbeta,\bsigma^2)$, say $(\hat{\balpha},\hat{\bbeta},\hat{\bsigma}^2)$, then $\bpi(\cdot)$ can be estimated locally by maximizing the following local log-likelihood function:
\begin{equation}
\ell^{(2)}_1(\bpi)=\sum_{i=1}^n\log\{\sum_{j=1}^k\pi_j(\hat{\balpha}^T\bx_i)\phi(Y_i|\bx_i^T\hat{\bbeta}_j,\hat{\sigma}^2_j)\}K_h(\hat{\balpha}^T\bx_i-z).
\label{p41-l1}
\end{equation}

Let $\hat{\bpi}(\cdot)$ be the estimate that maximizes (\ref{p41-l1}). We can then further update the estimate of $(\balpha,\bbeta,\bsigma^2)$ by maximizing
\begin{equation}
\ell^{(2)}_2(\balpha,\bbeta,\bsigma^2)=\sum_{i=1}^n\log\{\sum_{j=1}^k\hat{\pi}_j(\balpha^T\bx_i)\phi(Y_i|\bx_i^T\bbeta_j,\sigma^2_j)\}.
\label{p41-l2}
\end{equation}
The backfitting algorithm by iterating the above two steps can be summarized as
follows.
%The computing algorithm is very similar to Section \ref{sec:comp}, thus, is omitted here.
%\subsubsection{Computing Algorithm}

\begin{alg}
\label{backfittingalg}
Backfitting algorithm to estimate the model (\ref{p41-model}).
\begin{description}
\item[Step 1:] Obtain an initial estimate of $(\balpha,\bbeta,\bsigma^2)$.
\item[Step 2:] Given $(\hat{\balpha},\hat{\bbeta},\hat{\bsigma}^2)$, use the following modified EM-type algorithm to maximize $\ell^{(2)}_1$ in (\ref{p41-l1}).\\
\underline{\textbf{E-step:}} Calculate the expectations of component labels based on estimates from $l^{th}$ iteration:
\begin{equation}
p_{ij}^{(l+1)}=\frac{\pi^{(l)}_j(\hat{\balpha}^T\bx_i)\phi(Y_i|\bx_i^T\hat{\bbeta}_j,\hat{\sigma}_j^2)}
{\sum_{j=1}^k\pi^{(l)}_j(\hat{\balpha}^T\bx_i)\phi(Y_i|\bx_i^T\hat{\bbeta}_j,\hat{\sigma}_j^2)},
\end{equation}
where $i=1,\ldots,n, j=1,...,k$.
\underline{\textbf{M-step:}} Update the estimate
\begin{equation}
\pi_j^{(l+1)}(z)=\frac{\sum_{i=1}^np_{ij}^{(l+1)}K_h(\hat{\balpha}^T\bx_i-z)}{\sum_{i=1}^nK_h(\hat{\balpha}^T\bx_i-z)}
\end{equation}
for $z\in\{u_t,t=1,...,N\}$. We then update $\pi_j^{(l+1)}(\hat{\balpha}^T\bx_i)$, $i=1,...,n$ by linear interpolating $\pi_j^{(l+1)}(u_t)$, $t=1,...,N$.

\item[Step 3:] Given $\hat{\bpi}(\cdot)$ from Step 2, update $(\hat{\balpha},\hat{\bbeta},\hat{\bsigma}^2)$ by maximizing (\ref{p41-l2}). We propose to iterate between updating $\balpha$ and $(\bbeta, \bsigma)$.

\medskip

Step 3.1: Given $\hat{\balpha}$, update $(\bbeta,\bsigma^2)$.\\
\underline{\textbf{E-step:}} Calculate the classification probabilities:
\begin{equation}
p_{ij}^{(l+1)}=\frac{\hat{\pi}_j(\hat{\balpha}^T\bx_i)\phi(Y_i|\bx_i^T\bbeta^{(l)}_j,\sigma^{2(l)}_j)}
{\sum_{j=1}^k\hat{\pi}_j(\hat{\balpha}^T\bx_i)\phi(Y_i|\bx_i^T\bbeta^{(l)}_j,\sigma^{2(l)}_j)},\quad j=1,...,k.
\end{equation}
\underline{\textbf{M-step:}} Update $\bbeta$ and $\bsigma^2$:
\begin{align}
\bbeta_j^{(l+1)}&=(\bS^T\bR_j^{(l+1)}\bS)^{-1}\bS^T\bR_j^{(l+1)}\by,\\
\sigma_j^{2(l+1)}&=\frac{\sum_{i=1}^np_{ij}^{(l+1)}(Y_i-\bx_i^T\bbeta_j^{(l+1)})^2}{\sum_{i=1}^np_{ij}^{(l+1)}},
\end{align}
where $j=1,...,k$, $\bR_j^{(l+1)}=diag\{p_{ij}^{(l+1)},...,p_{nj}^{(l+1)}\}$, and $\bS=(\bx_1,...,\bx_n)^T$.

\smallskip

Step 3.2: Given $(\hat{\bbeta},\hat{\bsigma}^2)$, update $\balpha$ by maximizing the following log-likelihood \[\ell^{(2)}_3(\balpha)=\sum_{i=1}^n\log\{\sum_{j=1}^k\hat{\pi}_j(\balpha^T\bx_i)\phi(Y_i|\bx_i^T\hat{\bbeta}_j,\hat{\sigma}^2_j)\}.\]

\smallskip

Step 3.3: Iterate Steps 3.1-3.2 until convergence.
\item[Step 4:] Iterate Steps 2-3 until convergence.
\end{description}
\end{alg}
There are many ways to obtain an initial estimate of $(\balpha,\bbeta,\bsigma^2)$. In our numerical studies, we get an initial estimate of $(\bbeta,\bsigma^2)$ by fitting traditional mixtures of linear regression models. Using resulting hard-clustering results as new response variable, we apply SIR to get an initial estimate of $\balpha$.

\subsection{Asymptotic Properties}
Let $(\hat{\bpi}(z),\hat{\balpha},\hat{\bbeta},\hat{\bsigma}^2)$ be the resulting estimate of backfitting Algorithm \ref{backfittingalg}. In this section, we investigate their asymptotic properties. Let $\bta=(\bbeta^T,(\bsigma^2)^T)^T$ and $\blambda=(\balpha^T,\bta^T)^T$. Define
\begin{align*}
\ell(\bpi(z),\blambda,\bx,y)&=\log \sum_{j=1}^k\pi_j(z)\phi\{y|\bx^T\bbeta_j,\sigma_j^2\},\\
 q_{\pi}(z)&=\frac{\partial\ell(\bpi(z),\lambda,x,y)}{\partial\bpi},\\ q_{\pi\pi}(z)&=\frac{\partial^2\ell(\bpi(z),\lambda,x,y)}{\partial\bpi\partial\bpi^T}. \end{align*}
Similarly, define $q_{\lambda}$, $q_{\lambda\lambda}$, and $q_{\pi\eta}$.
Denote $\mathcal{I}^{(2)}_\pi(z)=-E[q_{\pi\pi}(Z)|Z=z]$ and $\Lambda_2(u|z)=E[q_{\pi}(z)|Z=u]$.

Under some regularity conditions, the asymptotic properties of $\hat{\bpi}(z)$ are given in the following theorem and its proof is given in the supplemental material.
\begin{thm}
\label{p4-thm:nonp}
Assume that conditions (C1)-(C4) and (C9)-(C11) in the supplemental material hold. Then, as $n\rightarrow\infty$, $h\rightarrow 0$ and $nh\rightarrow\infty$, we have
\begin{align}
\sqrt{nh}\{\hat{\bpi}(z)-\bpi(z)-\mathcal{B}_2(z)+o_p(h^2)\}\overset{D}{\rightarrow} N\{0,\nu_0f^{-1}(z)\mathcal{I}^{(2)}_\pi(z)\},
\end{align}
where \[\mathcal{B}_2(z)=\mathcal{I}_\pi^{(2)-1}\left\{\frac{f'(z)\Lambda_2'(z|z)}{f(z)}+\frac{1}{2}\Lambda_2''(z|z)\right\}\kappa_2h^2.\]
\end{thm}

The asymptotic property of the parametric estimate $\hat{\blambda}$ is given in the following theorem and its proof is given in the supplemental material.
\begin{thm}
\label{p4-thm:para}
Assume that conditions (C1)-(C4) and (C9)-(C12) in the supplemental material hold. Then, as $n\rightarrow\infty$, $nh^4\rightarrow 0$, and $nh^2/\log(1/h)\rightarrow\infty$,
\begin{equation}
\sqrt{n}(\hat{\blambda}-\blambda)\overset{D}{\rightarrow} N(0,\bQ_2^{-1}),\notag
\end{equation}
where,
\[\bQ_2=E\left[q_{\pi\pi}(Z)\begin{pmatrix}\bx\bpi'(Z)\\\textbf{I}\end{pmatrix}\left\{
\begin{pmatrix}\bx\bpi'(Z)\\\textbf{I}\end{pmatrix}-
\begin{pmatrix}\mathcal{I}_\pi^{(2)-1}(Z)E\{q_{\pi\pi}(Z)(\bx\bpi'(Z))^T|Z\}\\
\mathcal{I}_\pi^{(2)-1}(Z)E\{q_{\pi\eta}(Z)|Z\}\end{pmatrix}\right\}^T\right].\]
\end{thm}

\section{Simulation Studies}
\label{sec:p3-s4}
%\subsection{Simulation Study}
In this section, we conduct simulation studies to test the performance of the proposed models and estimation procedures.

The performance of the estimates of the mean functions $m_j(\cdot)$'s in the model (\ref{p3-model}) is measured by the square root of the average square errors (RASE)
\begin{equation}
RASE_m^2=N^{-1}\sum_{j=1}^k\sum_{t=1}^N[\hat{m}_j(u_t)-m_j(u_t)]^2.\notag
\end{equation}
In our simulation, we set $N=100$. Similarly, we can define the $RASE$ for variance functions $\sigma_j^2(\cdot)$'s and proportion functions $\pi_j(\cdot)$'s, denoted by $RASE_{\sigma^2}$ and $RASE_\pi$, respectively.

\medskip

%To apply the proposed methodologies, we use cross-validation (CV) to select a proper bandwidth for estimating the nonparametric functions.

\emph{Example 1:} We conduct a simulation for a two-component MSIM:
\begin{center}
$\pi_1(z)=0.5+0.3\sin(\pi z)$ and $\pi_2(z)=1-\pi_1(z)$,\\
$m_1(z)=3-\sin(2\pi z/\sqrt{3})$ and $m_2(z)=\cos(\sqrt{3}\pi z)$,\\
$\sigma_1(z)=0.7+\sin(3\pi z)/15$ and $\sigma_2(z)=0.3+\cos(1.3\pi z)/10$.
\end{center}
where $z_i=\balpha^T\bx_i$, $\bx_i$ are trivariate with independent uniform (0,1) components, and the direction parameter is $\balpha=(1,1,1)/\sqrt{3}$. The sample sizes $n=200$, $n=400$, and $n=800$ are conducted over $500$ repetitions. To estimate $\balpha$, we use sliced inverse regression (SIR) and the fully iterative backfitting estimate (FIB). To estimate the nonparametric functions, we apply the one-step estimate (OS) and FIB. For FIB, we use both true value (T) and SIR (S) as the initial values.

We first select a proper bandwidth for estimating $\bpi(\cdot)$, $\bm(\cdot)$ and $\bsigma^2(\cdot)$. Based on Theorem \ref{p3-thm:nonp}, one can calculate theoretical optimal bandwidth by minimizing asymptotic mean squared errors. However, the theoretical optimal bandwidth depends on many unknown quantities, which are not easy to estimate in practice. In our examples, we propose to use the following cross-validation (CV) method to choose the bandwidth. Let $\mathscr{D}$ be the full data set, and divide $\mathscr{D}$ into a training set $\mathscr{R}_l$ and a test set $\mathscr{T}_l$. That is, $\mathscr{R}_l\cup \mathscr{T}_l=\mathscr{D}$ for $l=1,...,L$. We use the training set $\mathscr{R}_l$ to obtain the estimates $\{\hat{\bpi}(\cdot),\hat{\bm}(\cdot),\hat{\bsigma}^2(\cdot),\hat{\balpha}\}$. We then evaluate $\bpi(\cdot)$, $\bm(\cdot)$ and $\bsigma^2(\cdot)$ for the test data set. For each $(\bx_t,y_t)\in\mathscr{T}_l$, we calculate the classification probability as
\begin{equation}
\hat{p}_{tj}=\frac{\hat{\pi}_j(\hat{\balpha}^T\bx_t)\phi(y_t|\hat{m}_j(\hat{\balpha}^T\bx_t),\hat{\sigma}_j^2(\hat{\balpha}^T\bx_t))}{\sum_{j=1}^k\hat{\pi}_j(\hat{\balpha}^T\bx_t)\phi(y_t|\hat{m}_j(\hat{\balpha}^T\bx_t),\hat{\sigma}_j^2(\hat{\balpha}^T\bx_t))},
\end{equation}
for $j=1,...,k$. We consider the regular $CV$, which is defined by
\begin{equation}
CV(h)=\sum_{l=1}^L\sum_{t\in \mathscr{T}_l}(y_t-\hat{y}_t)^2\notag,
\end{equation}
where $\hat{y}_t=\sum_{j=1}^k\hat{p}_{tj}\hat{m}_j(\hat{\balpha}^T\bx_t)$. We also implemented the likelihood based cross validation to choose the bandwidth and the results are similar but with more computations.

We set $L=10$ and randomly partition the data. We repeat the procedure 30 times, and take the average of the selected bandwidth as the optimal bandwidth, denoted by $\hat{h}$. In the simulation, we consider three different bandwidths, $\hat{h}\times n^{-2/15}$, $\hat{h}$ and $1.5\hat{h}$, which correspond to the under-smoothing, appropriate smoothing and over-smoothing condition, respectively.

Table \ref{p3-table1} reports the MSEs of $\hat{\balpha}$ (true value times 100) and Table \ref{p3-table2} contains the mean and standard deviation of $RASE_\pi$, $RASE_m$, and $RASE_{\sigma^2}$. Based on Table \ref{p3-table1}, we can see that the proposed fully iterative backfitting estimates (FIB) give much better results than SIR, which is reasonable since FIB makes use of mixture information while SIR does not. Based on Table \ref{p3-table2}, we can see that OS provides close estimates to FIB, although FIB generally provides slightly smaller RASEs than OS for finite sample size. This verified the theoretical results stated in Section \ref{sec:asympmodel1}.

In addition, from Tables \ref{p3-table1} and \ref{p3-table2}, we can see that the proposed bandwidth selection procedure based on cross validation works reasonably well since the appropriate bandwidths chosen by CV usually provide the estimate that is or is close to the best one. Furthermore, FIB(S) provides similar results to FIB(T). Therefore, SIR provides good initial values for the proposed fully iterative estimates.

% We see that the fully iterative estimate is not sensitive to initial values.%the fully iterative estimates provide better results than one-step estimate under appropriate-smoothing condition, and the fully iterative estimate is not sensitive to initial values.

\begin{table}[htb]
%\small
\centering
 %  \begin{center}
\caption{MSE of $\hat{\balpha}$ (true value times 100) for Example 1.} \vskip 0.05in
\def\arraystretch{1.3}
\small \hspace*{-22.75pt}
\begin{tabular}{ c c|c| c c c | c c cc } \hline
\hline
&&SIR&&FIB(T)&&&FIB(S)&\\
\hline
&&&$h=0.054$&$h=0.109$&$h=0.164$&$h=0.054$&$h=0.109$&$h=0.164$\\
&$\alpha_1$&0.881&0.099&0.126&0.128&0.287&0.130&0.147\\
$n=200$&$\alpha_2$&0.829&0.113&0.144&0.124&0.324&0.144&0.137\\
&$\alpha_3$&1.066&0.110&0.152&0.137&0.388&0.154&0.167\\
\hline
&&&$h=0.045$&$h=0.100$&$h=0.149$&$h=0.045$&$h=0.100$&$h=0.149$\\
&$\alpha_1$&0.435&0.066&0.046&0.046&0.125&0.050&0.045\\
$n=400$&$\alpha_2$&0.447&0.063&0.054&0.051&0.121&0.055&0.052\\
&$\alpha_3$&0.411&0.062&0.052&0.052&0.123&0.053&0.052\\
\hline
&&&$h=0.037$&$h=0.091$&$h=0.137$&$h=0.037$&$h=0.091$&$h=0.137$\\
&$\alpha_1$&0.215&0.047&0.022&0.029&0.063&0.035&0.024\\
$n=800$&$\alpha_2$&0.256&0.034&0.035&0.040&0.044&0.029&0.027\\
&$\alpha_3$&0.226&0.065&0.031&0.058&0.062&0.050&0.030\\
\hline
\end{tabular}
%\end{center}
\label{p3-table1}
\end{table}

\begin{table}[htb]
    %\small
 %  \begin{center}
 \centering
\caption{Mean and Standard Deviation of RASEs for Example 1.} \vskip 0.05in
\def\arraystretch{1.3}
\small \hspace*{-22.75pt}
\begin{tabular}{  c|c| c c c | c c cc } \hline
\hline
&OS&&FIB(T)&&&FIB(S)&\\
\hline
n=200&$h=0.125$&$h=0.054$&$h=0.109$&$h=0.164$&$h=0.054$&$h=0.109$&$h=0.164$\\
$\pi$& 0.044(0.017)&0.057(0.015) &0.043(0.016)&0.049(0.017)&0.058(0.015) &0.043(0.016) &0.049(0.017)\\
$\mu$&0.227(0.063) &0.181(0.098) &0.176(0.046)&0.287(0.056)&0.178(0.086) &0.177(0.051) &0.288(0.059)\\
$\sigma^2$&0.197(0.084) &0.175(0.169)&0.163(0.081)&0.246(0.071)&0.162(0.131)&0.164(0.095)&0.247(0.080)\\
\hline
n=400&$h=0.108$&$h=0.045$&$h=0.100$&$h=0.149$&$h=0.045$&$h=0.100$&$h=0.149$\\
$\pi$&0.023(0.008)&0.032(0.008) &0.023(0.008)&0.027(0.009)&0.032(0.008) & 0.023(0.008) &0.027(0.009)\\
$\mu$&0.118(0.022)&0.093(0.045)&0.100(0.022) &0.169(0.020)&0.094(0.046) &0.100(0.022)&0.169(0.020)\\
$\sigma^2$&0.104(0.035)& 0.089(0.077)&0.093(0.045)&0.143(0.028)&0.089(0.077)&0.093(0.045)&0.143(0.028)\\
\hline
n=800&$h=0.094$&$h=0.037$&$h=0.091$&$h=0.137$&$h=0.037$&$h=0.091$&$h=0.137$\\
$\pi$&0.013(0.004)&0.017(0.003)&0.012(0.004)&0.016(0.004)&0.017(0.003)&0.012(0.004)&0.016(0.004)\\
$\mu$&0.062(0.010)&0.050(0.023)&0.056(0.010)&0.102(0.011)&0.050(0.023)&0.056(0.010)&0.101(0.010)\\
$\sigma^2$&0.055(0.015)&0.049(0.046)&0.052(0.015)&0.086(0.010)&0.049(0.046)&0.051(0.012)&0.085(0.010)\\
\hline
\end{tabular}
%\end{center}
\label{p3-table2}
\end{table}

%\begin{figure}
%\vspace{6pc}
%\includegraphics[height=3in,width=5in]{p3-n400.pdf}
%\includegraphics[scale=0.7]{p3-n400.pdf}
%\caption[]{Simulation results of  for $n=800$ and $h=0.091$. The FIB(S) (dashed line) and true function (solid line) of: (a) $\pi_1$; (b) $\mu_1$ and $\mu_2$; (c) $\sigma_1^2$; and (d) $\sigma_2^2$.}
%\label{p3-figure2}
%\end{figure}

\medskip

\emph{Example 2:} We conduct a simulation for a two-component MRSIP:
\begin{center}
$\pi_1(z)=0.5-0.35\sin(\pi z)$ and $\pi_2(z)=1-\pi_1(z)$,\\
$m_1(\bx)=1+3x_2$ and $m_2(\bx)=-1+2x_1+3x_3$,\\
$\sigma^2_1=0.7$ and $\sigma^2_2=0.6$,
\end{center}
where $m_1(\bx)$ and $m_2(\bx)$ are the regression functions for the first and second components, respectively. Therefore, $\bbeta_1=(1,0,3,0)$ and $\bbeta_2=(-1,2,0,3).$ $\bx_i$ are trivariate with independent uniform (0,1) components, and the single index parameter is $\balpha=(1,1,1)/\sqrt{3}$. MRSIP with true value (T) and SIR (S) as initial values are used to fit the data, and the results are compared to the traditional mixture of linear regression models (MixLinReg). The bandwidth for MRSIP is chosen based on the cross validation similar to Example 1.

Table \ref{p41-tab1} reports the MSEs of parameter estimates, and Table \ref{p41-tab2} contains the MSEs of $\hat{\balpha}$ and the average of $RASE_\pi$. From Table \ref{p41-tab1}, we can see that MRSIP works comparable to MixLinReg when the sample size is small, and outperforms MixLinReg when sample size is large (such as $n=400$ or $800$). By reducing the modelling bias of component proportions, MRSIP is able to better classify observations into two components and thus provide better component regression parameters. Based on Table \ref{p41-tab2}, it is clear that MRSIP provides better estimates of component proportions than MixLinReg since the constant assumption of component proportions by MixLinReg is violated. From both tables, we can see that MRSIP(S) provides similar results to MRSIP(T), which demonstrates that SIR provides good initial values for MRSIP.

\begin{table}[htb]
    \centering
\caption{The MSEs of parameters (true value times 100) for Example 2.} \vskip 0.05in
\def\arraystretch{1.3}
\small \hspace*{-22.75pt}
\begin{tabular}{c c| c c c c c c c c | c cc } \hline
& & $\beta_{10}$ & $\beta_{11}$ & $\beta_{12}$ & $\beta_{13}$ & $\beta_{20}$ & $\beta_{21}$ & $\beta_{22}$ & $\beta_{23}$ & $\sigma^2_{1}$ & $\sigma^2_{2}$\\\hline
$n=200$ & MRSIP(S) & 46.37&32.78&34.73&37.61&11.19&16.55&15.05&16.36&4.649&1.754\\
\cline{2-12}
&MRSIP(T)& 51.91&33.62&39.01&37.25&11.10&16.56&15.07&16.04&4.584&1.649\\
\cline{2-12}
$h=0.131$& MixLinReg & 50.87&33.67&42.53&34.68&12.03&12.66&18.84&12.30& 4.250&1.265\\
\hline
$n=400$ & MRSIP(S) & 13.83&11.89&14.19&11.47&5.541&6.332&6.767&7.165&1.631&0.721\\
\cline{2-12}
&MRSIP(T)& 14.79&12.49&14.84&11.59&5.513&6.254&6.632&6.926&1.672&0.675\\
\cline{2-12}
$h=0.103$& MixLinReg & 29.03&14.97&29.46&15.72&8.045&5.967&12.46&6.269&1.864&0.626\\
\hline
$n=800$ & MRSIP(S) &  6.324&4.491&6.150&4.736&2.365&2.973&2.773&3.584&0.669&0.334\\
\cline{2-12}
&MRSIP(T)& 6.788&4.614&6.820&4.922&2.301&2.829&2.718&3.348&0.691&0.307\\
\cline{2-12}
$h=0.080$& MixLinReg & 21.89&6.866&21.84&8.223&5.413&3.163&8.775&3.640&0.848&0.352\\
\hline
\end{tabular}
\label{p41-tab1}
\end{table}

\begin{table}[htb]
    \centering
\caption{The MSEs of single index parameter and the average of RASE$_\pi$ (true value times 100) for Example 2.} \vskip 0.05in
\def\arraystretch{1.3}
\small \hspace*{-22.75pt}
\begin{tabular}{c c | c c c | cc } \hline
&&$\alpha_1$&$\alpha_2$&$\alpha_3$&RASE$_\pi$\\\hline
$n=200$&MRSIP(S)&  5.709&19.30&5.996&18.87\\
\cline{2-7}
&MRSIP(T)&   4.984&9.449&4.896&17.86\\
\cline{2-7}
$h=0.131$&MixLinReg&-&-&-&   28.98\\
\hline
$n=400$&MRSIP(S)&   2.682&6.968&3.029&13.74\\
\cline{2-7}
&MRSIP(T)&   2.113&3.019&1.902&12.98\\
\cline{2-7}
$h=0.103$&MixLinReg&-&-&-&    28.23\\
\hline
$n=800$&MRSIP(S)&   0.980&2.527&1.585&10.35\\
\cline{2-7}
&MRSIP(T)&    0.892&0.979&0.969&9.960\\
\cline{2-7}
$h=0.080$&MixLinReg&-&-&-&     28.04&\\
\hline
\end{tabular}
\label{p41-tab2}
\end{table}

\section{Real Data Example}
\label{realdata}
We illustrate the proposed methodology by an analysis of ``The effectiveness of National Basketball Association guards''.
There are many ways to measure the (statistical) performance of guards in the National Basket Association (NBA). Of interest is how the height of the player (Height), minutes per game (MPG) and free throw percentage (FTP) affect points per game (PPM) \citep{chatterjee1995}.

\begin{figure}[h!]
\begin{center}
\scalebox{0.45}{\includegraphics{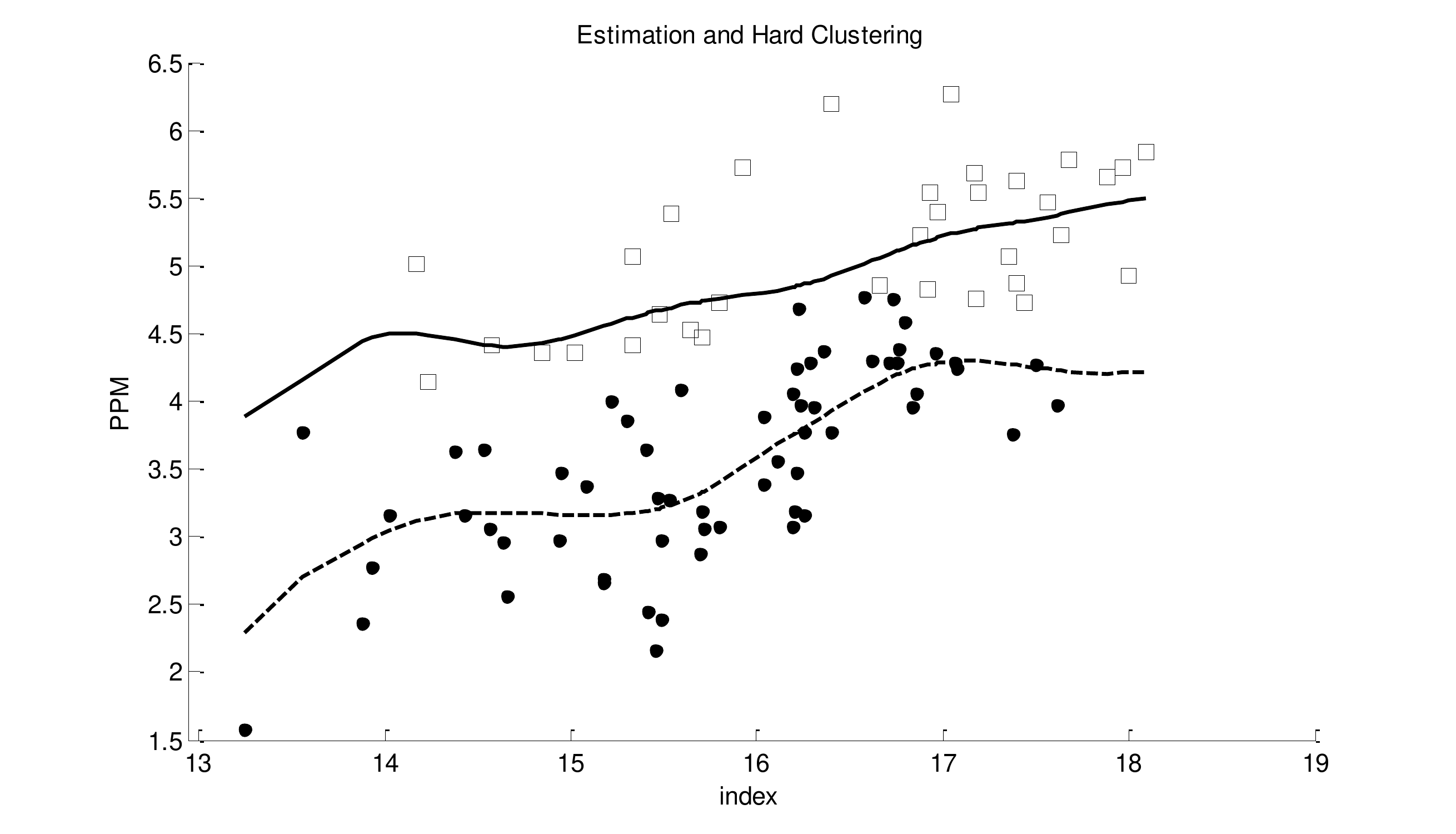}}
 \put(-120,-15){(a)}\\
\scalebox{0.45}{\includegraphics{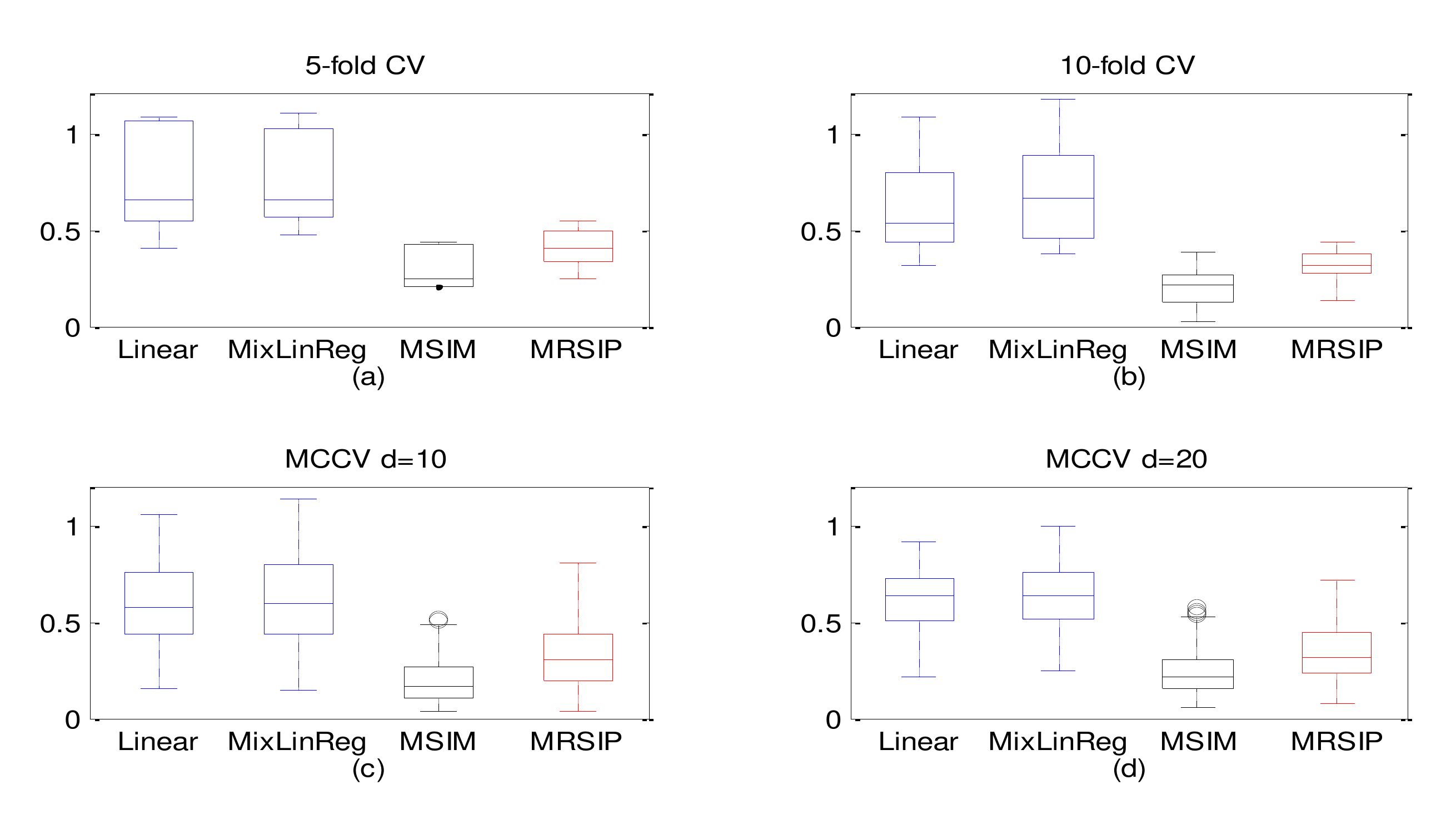}}
 \put(-120,-15){(b)}
\end{center}
\caption[]{NBA data: (a) Estimated mean functions and a hard-clustering result; (b) Prediction accuracy: 5-fold CV; 10-fold CV; MCCV d=10; MCCV d=20.}
\label{p3-figure3}
\end{figure}

The data set contains some descriptive statistics for all 105 guards for the 1992-1993 season. Since players playing very few minutes are quite different from those who play a sizable part of the season, we only look at those players playing 10 or more minutes per game and appearing in 10 or more games. In addition, Michael Jordan is an outlier, so we also omit him from our data analysis. These exclude 10 players  (Chatterjee et al., 1995). We divide each variable by its corresponding standard deviation, so that they have comparable numerical scales. An optimal bandwidth is selected at 0.344 by CV procedure. Figure \ref{p3-figure3}(a) contains the estimated mean functions and hard-clustering results, denoted by dots and squares, respectively. The 95\% confidence interval for $\hat{\balpha}$ based on MSIM are (0.134,0.541), (0.715,0.949) and (0.202,0.679). Therefore, MPG is the most influential factor on PPM. This might be partly explained by that coaches tend to let good players with higher PPM play longer minutes per game (i.e., higher MPG).

To evaluate the prediction performance of the proposed models and compared them to linear regression model and mixture of linear regression models, we used $d$-fold cross-validation with $d$=5, 10, and also Monte-Carlo cross-validation (MCCV) \citep{shao1993}. In MCCV, the data were partitioned 500 times into disjoint training subsets (with size $n-d$) and test subsets (with size $d$). The mean squared prediction error evaluated at the test data sets over 500 replications are reported as boxplots in Figure \ref{p3-figure3}(b). Apparently, the MSIM and the MRSIP have superior prediction power than the linear regression model or the mixture of linear regression models, and MSIM is more favorable than the MRSIP for this data set. The two groups of guards our new models found might be explained by the difference between shooting guards and passing guards.

\section{Discussion}
\label{discussion}
In this paper, we propose two finite semiparametric mixture of regression models and provide the modified EM algorithms to estimate them. We establish the identifiability results of the new models and investigate the asymptotic properties of the proposed estimation procedures. Throughout the article, we assume that the number of components is known and fixed, but it requires more research to select the number of components for the proposed semiparametric mixture models. It will be interesting to know whether the recently proposed EM test \citep{chen09,li10} can be extended to the proposed semiparametric mixture models. In addition, it is also interesting to build some formal model selection procedure to compare different semiparametric mixture models. In the real data application, we use the cross-validation criteria to compare different models. When the models are nested, one might use generalized likelihood ratio statistic proposed by \citet{fan01} to test any parametric assumption for the semiparametric models. Furthermore, the assumption of fixed dimension of predictors can be relaxed and the proposed models can be extended to the cases where the dimension of predictors $p$ also diverges with the sample size $n$. This might be done by using the idea of penalized local likelihood if the sparsity assumption is added on the predictors.

\end{document}